\documentclass[12pt]{iopart}
\usepackage{cite}
\usepackage[dvips]{graphicx}

\let\Otemize =\itemize
\let\Onumerate =\enumerate
\let\Oescription =\description
\def\Nospacing{\itemsep=0pt\topsep=0pt\partopsep=0pt\parskip=0pt\parsep=0pt}
\def\Topspac{\vspace{-0.5\baselineskip}}
\def\Botspac{\vspace{-0.2\baselineskip}}
\newenvironment{Itemize}{\Topspac\Otemize\Nospacing}{\endlist\Botspac}

\renewcommand{\lsim}{~{\buildrel < \over {_\sim}}~}
\newcommand{\gsim}{~{\buildrel > \over {_\sim}}~}
\newcommand{\sqrtsNN}{\sqrt{s_{\scriptscriptstyle{{\rm NN}}}}}
\newcommand{\av}[1]{\left\langle #1 \right\rangle}

\newcommand{\mev}{\mathrm{MeV}}
\newcommand{\gev}{\mathrm{GeV}}
\newcommand{\tev}{\mathrm{TeV}}
\newcommand{\fm}{\mathrm{fm}}

\newcommand{\cm}{\mathrm{cm}}

\newcommand{\mum}{\mathrm{\mu m}}

\newcommand{\RAA}{R_{\rm AA}}

\renewcommand{\pt}{p_{\rm t}}
\renewcommand{\d}{{\rm d}}
\newcommand{\dEdx}{{\rm d}E/{\rm d}x}
\newcommand{\dNdy}{{\rm d}N_{\rm ch}/{\rm d}y}

\newcommand{\ccbar}{\mbox{$\mathrm {c\overline{c}}$}}
\newcommand{\bbbar}{\mbox{$\mathrm {b\overline{b}}$}}

\newcommand{\Dz}{\mbox{$\mathrm {D^0}$}}

\begin{document}

\title{ALICE potential for heavy-flavour physics}

\author{A Dainese}

\address{INFN - Laboratori Nazionali di Legnaro, 
         viale dell'Universit\`a 2, 35020 Legnaro (Padova), Italy. e-mail: andrea.dainese@lnl.infn.it}

\begin{abstract}
The Large Hadron Collider (LHC), where lead nuclei will collide
at the unprecedented c.m.s. energy of 5.5 TeV per nucleon--nucleon pair, 
will offer new and unique opportunities for the study of the properties of 
strongly interacting matter at high energy density over extended volumes.
We will briefly explain why heavy-flavour particles are well-suited tools
for such a study and we will describe how the ALICE experiment is preparing 
to make use of these tools.
\end{abstract}


\section{Introduction}
\label{intro}

ALICE~\cite{alicePPR1,alicePPR2} is the dedicated heavy-ion experiment at the Large
Hadron Collider (LHC). The main physics goal of the experiment is the 
study of strongly-interacting matter in the conditions of high energy 
density ($>10~\gev/\fm^3$) and high temperature ($\gsim 0.2~\gev$),
expected to be reached in central \mbox{Pb--Pb} collisions at 
$\sqrtsNN=5.5~\tev$. 
Under these 
conditions, according to lattice QCD calculations, quark confinement into 
colourless hadrons should be removed and
a deconfined Quark--Gluon Plasma should be formed~\cite{alicePPR1}.

As we will detail in the next section, heavy-flavour particles
are regarded as effective probes of the system conditions. 
In particular: 
\begin{Itemize}
\item open charm and beauty hadrons would be sensitive to the energy density,
through the mechanism of in-medium energy loss of heavy quarks;
\item quarkonium states would be sensitive to the initial temperature of the
system through their dissociation due to colour screening.
\end{Itemize}

\section{Heavy-quark and quarkonia phenomenology at the LHC}
\label{pheno}

\begin{table}[!t]
\caption{Expected $\rm Q\overline Q$ yields at the LHC, 
        from NLO pQCD~\cite{notehvq}.
         For \mbox{Pb--Pb}, 
         nuclear shadowing of the parton distribution functions 
         is included and binary scaling is applied.}
\label{tab:xsec}
\begin{center}
\begin{tabular}{ccccc}
\hline
colliding system & $\sqrtsNN$ & centrality & $N^{\rm c\overline{c}}$/event  &  $N^{\rm b\overline{b}}$/event  \\
\hline
pp & $14~\tev$ & minimum bias  &  0.16 & 0.0072 \\
\mbox{Pb--Pb} & $5.5~\tev$ & central (0--5\% $\sigma^{\rm tot}$) & 115 & 4.6 \\
\hline
\end{tabular}
\end{center}
\end{table}

The expected yields for heavy-quark production 
in pp collisions at $\sqrt{s}=14~\tev$
are reported in the first line of Table~\ref{tab:xsec}~\cite{notehvq},
as obtained from the next-to-leading order perturbative QCD calculation  
implemented in 
the MNR program~\cite{mnr} with {\it reasonable} parameter values 
(quark masses and perturbative scales)~\cite{notehvq}. 
These yields have large uncertainties, of about a factor 2,
estimated by varying the parameters~\cite{notehvq}. 
In the table we report also the $\ccbar$ and $\bbbar$ yields 
in central \mbox{Pb--Pb} collisions at $\sqrtsNN=5.5~\tev$, 
obtained including in the
calculation the nuclear
modification of the parton distribution functions
(EKS98 parametrization~\cite{eks})
and applying a scaling of the hard yields 
with the mean number $\av{N_{\rm coll}}$ of binary nucleon--nucleon collisions 
in the \mbox{Pb--Pb} collision.

Experiments at the Relativistic Heavy Ion Collider (RHIC)
 have shown that the nuclear modification factor of 
particles $\pt$ distributions, 
\mbox{$R_{\rm AA}(\pt,\eta)=
{1\over \av{N_{\rm coll}}} \cdot 
{\d^2 N_{\rm AA}/\d\pt\d\eta \over 
\d^2 N_{\rm pp}/\d\pt\d\eta}$},
is a sensitive observable 
for the study of the interaction of the hard partons 
with the medium produced in nucleus--nucleus collisions.
Heavy quark medium-induced quenching is one of the most captivating 
topics to be 
addressed in \mbox{Pb--Pb} collisions at the LHC. Due to the 
QCD nature of parton energy loss, quarks are predicted to lose less
energy than gluons (that have a higher colour charge) and, in addition, 
the `dead-cone effect' is expected to reduce the energy loss of massive 
quarks with respect to light partons~\cite{dk,asw}. 
Therefore, one should observe a pattern 
of gradually decreasing $\RAA$ suppression when going from the mostly 
gluon-originated
light-flavour hadrons ($h^\pm$ or $\pi^0$) 
to D and to B mesons~\cite{adsw}: 
$\RAA^h\lsim\RAA^{\rm D}\lsim\RAA^{\rm B}$. The enhancement, with respect to 
unity, of {\it heavy-to-light $\RAA$ ratios}
has been suggested~\cite{adsw} as a sensitive observable to test the 
colour-charge ($R_{{\rm D}/h}=\RAA^{\rm D}/\RAA^h$) and mass 
($R_{{\rm B}/h}=\RAA^{\rm B}/\RAA^h$) dependence of parton energy loss.
In addition, the smaller predicted energy loss of b quarks 
may allow to overcome the main limitation of light partons 
as probes of the medium, namely their low sensitivity to the 
density of the medium. According to the theoretical 
models~\cite{fragility} based on the BDMPS framework~\cite{bdmps}, 
this so-called 
`fragility' became apparent already in \mbox{Au--Au} collisions at 
RHIC, and would be more pronounced in 
\mbox{Pb--Pb} collisions at the LHC, where hard gluons would either
be absorbed in the medium or escape the medium without really probing 
its density. Beauty quarks could instead provide stringent constraints on the 
energy density of the medium.

The measurement of D and B meson production cross sections will 
also serve as a baseline for the study of medium effects on quarkonia.
Two of the most interesting items in the quarkonia sector at the LHC 
will be: (a) understanding the 
interplay between colour-screening-induced 
suppression and statistical regeneration for 
J/$\psi$ production in a medium containing on the order of 100 $\ccbar$ pairs
(see Table~\ref{tab:xsec}); 
(b) measuring for the first time 
medium effects on the bottomonia resonances, expected to be available
with sufficient yields at the LHC. On this point, the predicted suppression 
pattern as a function of the plasma temperature is particularly
interesting: 
the $\Upsilon$ would melt at 
$T\approx 0.42~\gev$,
a temperature that would be reached only at the LHC, while the 
$\Upsilon^\prime$ would melt at the same temperature as the J/$\psi$,
$T\approx 0.19~\gev$. It will thus be important for the experiments to be able 
to measure also the $\Upsilon^\prime$, because, at variance with the 
J/$\psi$, it is expected to have a small regeneration probability and it would 
be very useful to disentangle J/$\psi$ suppression and regeneration.
In summary, measuring the in-medium dissociation probability of the different
quarkonium states at the LHC will provide an estimate of the initial 
medium temperature. 
We note that, in order to study the medium effects on charmonia, 
it will be mandatory to measure the fraction of secondary charmonia from B 
decays, expected to be about 20\% for the J/$\psi$,
in absence of medium-induced effects.

\section{Heavy-flavour detection in ALICE}
\label{exp}

The desing of the ALICE apparatus~\cite{alicePPR1} 
will allow the detection
of open heavy-flavour hadrons and quarkonia 
in the high-multiplicity environment 
of central \mbox{Pb--Pb} collisions at LHC energy, where up to few thousand 
charged particles might be produced per unit of rapidity. 
The heavy-flavour capability of the ALICE detector is provided by:
\begin{Itemize}
\item Tracking system; the Inner Tracking System (ITS), 
the Time Projection Chamber (TPC) and the Transition Radiation Detector (TRD),
embedded in a magnetic field of $0.5$~T, allow track reconstruction in 
the pseudorapidity range $|\eta|<0.9$ 
with a momentum resolution better than
2\% for $\pt<20~\gev/c$ 
and a transverse impact parameter\footnote{The transverse impact parameter,
$d_0$, is defined as the distance of closest approach of the track to the 
interaction vertex, in the plane transverse to the beam direction.} 
resolution better than 
$60~\mum$ for $\pt>1~\gev/c$ 
(the two innermost layers of the ITS are equipped with silicon pixel 
detectors).
\item Particle identification; charged hadrons ($\pi$, K, p) are separated via 
$\dEdx$ in the TPC and in the ITS and via time-of-flight measurement in the 
Time Of Flight (TOF) detector; electrons are separated from charged 
pions in the dedicated
Transition Radiation Detector (TRD), and in the TPC; 
muons are identified in the forward muon 
spectrometer covering in acceptance the range $-4<\eta<-2.5$. 
\end{Itemize}

Simulation studies~\cite{alicePPR2}
have shown that ALICE has good potential to carry out
a rich heavy-flavour physics programme. The main analyses in preparation 
are:
\begin{Itemize}
\item Open charm (section~\ref{open}): fully reconstructed hadronic decays 
$\rm D^0 \to K^-\pi^+$, $\rm D^+ \to K^-\pi^+\pi^+$,
$\rm D_s^+ \to K^-K^+\pi^+$ (under study), $\rm \Lambda_c^+ \to p K^-\pi^+$ (under study) in $|\eta|<0.9$.
\item Open beauty (section~\ref{open} and Ref.~\cite{julien}): 
inclusive single leptons ${\rm B\to e}+X$ 
in $|\eta|<0.9$ and ${\rm B\to\mu}+X$ in $-4<\eta<-2.5$; inclusive displaced
charmonia ${\rm B\to J/\psi\,(\to e^+e^-)}+X$ (under study).
\item Quarkonia (section~\ref{quarkonia}): $\psi$ and $\Upsilon$ states 
in the ${\rm e^+e^-}$ ($|\eta|<0.9$) and $\mu^+\mu^-$ ($-4<\eta<-2.5$) 
channels.
\end{Itemize} 
For all studies, a multiplicity of $\dNdy=4000$--$6000$
was assumed for central \mbox{Pb--Pb} collisions.
In the following, we report the results corresponding to the 
expected statistics collected by ALICE per LHC year: 
$10^7$ central (0--5\% $\sigma^{\rm inel}$) \mbox{Pb--Pb} events at
$\mathcal{L}_{\rm Pb-Pb}=10^{27}~\cm^{-2}{\rm s}^{-1}$
and $10^9$ pp events at 
$\mathcal{L}_{\rm pp}^{\rm ALICE}=5\times 10^{30}~\cm^{-2}{\rm s}^{-1}$,
in the barrel detectors; the forward muon arm will collect
about 40 times larger samples (i.e.\, $4\times 10^8$ central \mbox{Pb--Pb} events).

\section{Open charm and beauty capabilities}
\label{open}

\begin{figure}[!t]
  \begin{center}
    \includegraphics[width=.44\textwidth]{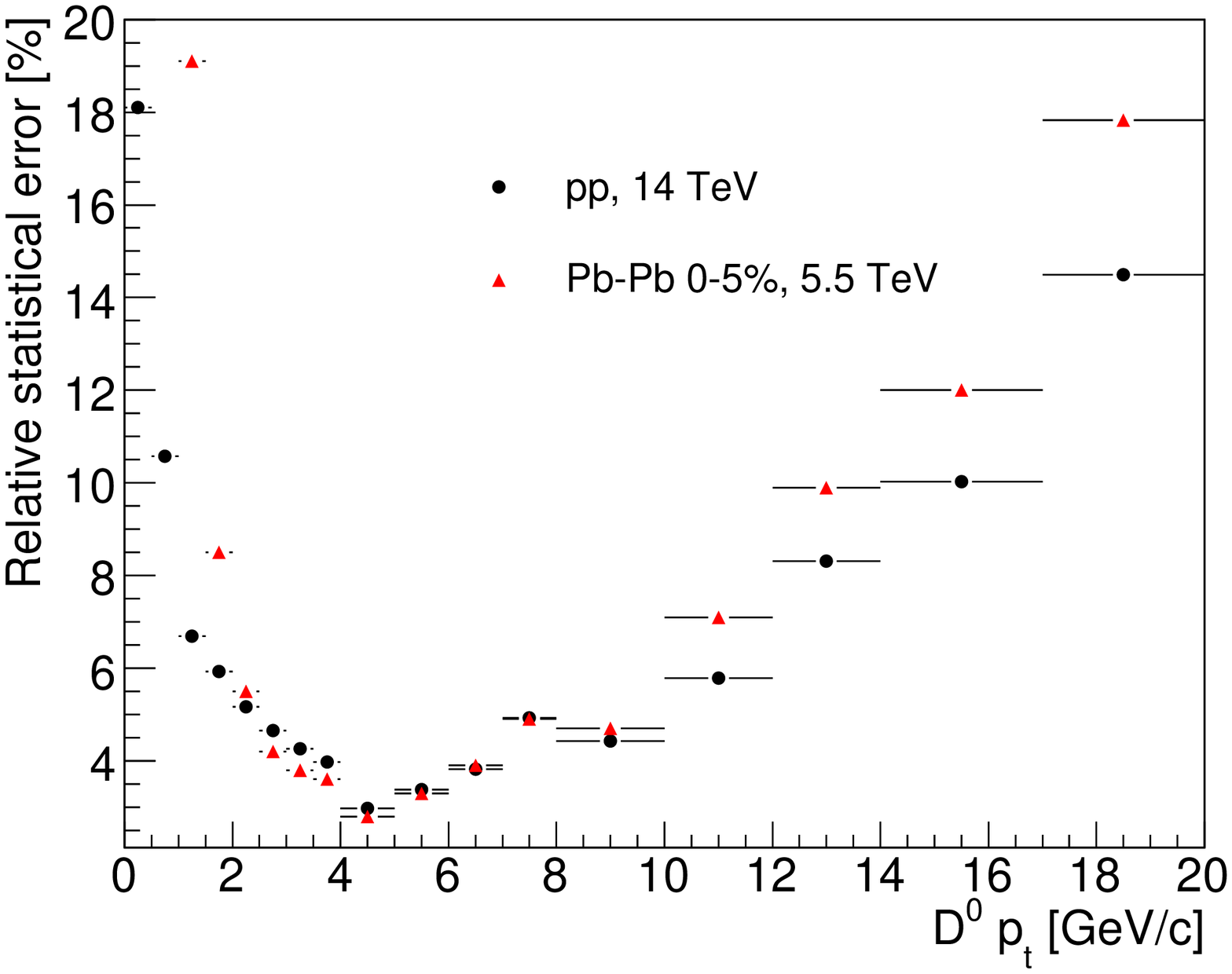}
    \includegraphics[width=.55\textwidth]{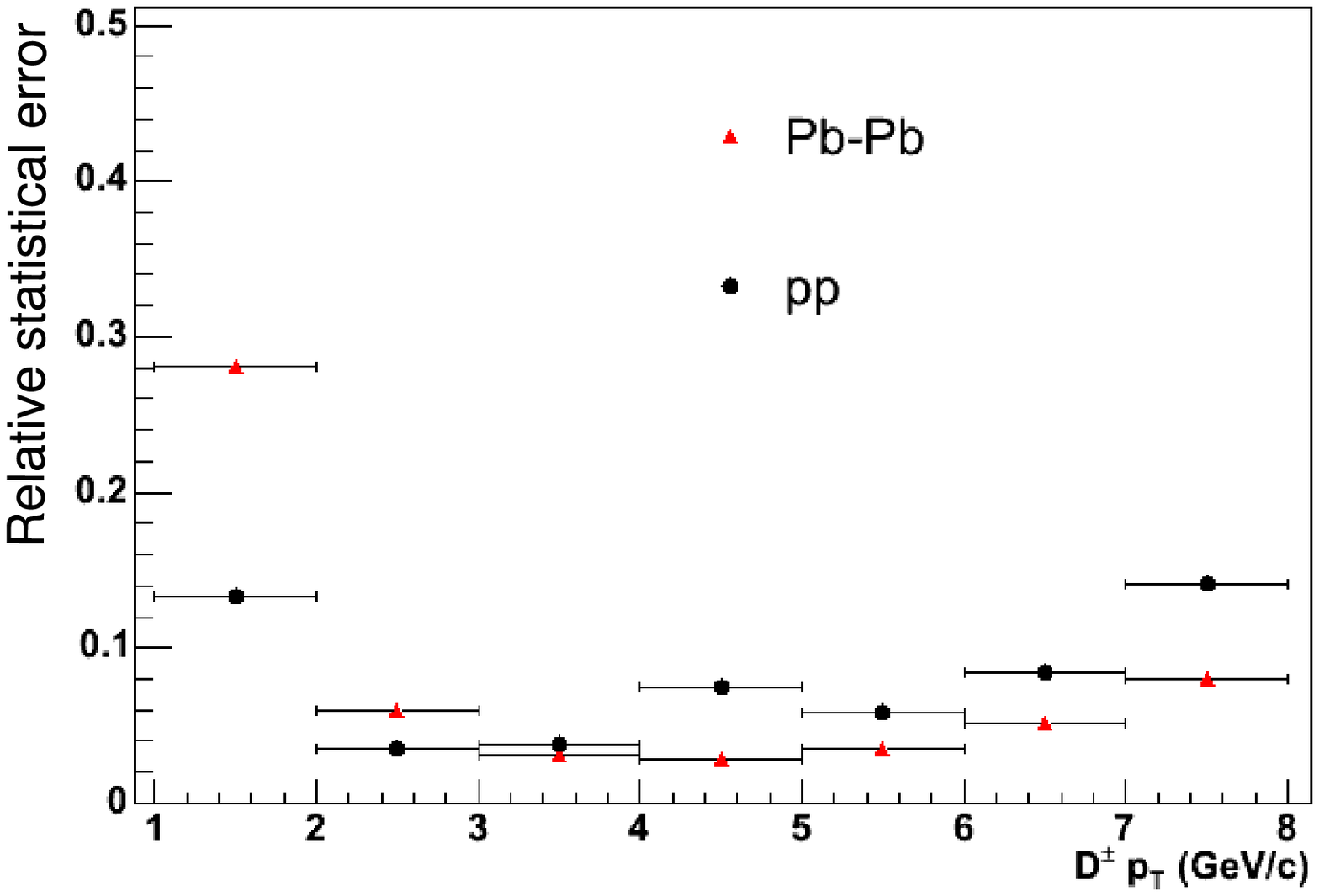}
    \caption{Expected relative statistical errors for the measurement 
             in ALICE
             of the production cross sections of ${\rm D^0}$ in the 
             ${\rm K^-\pi^+}$ channel (left) and ${\rm D^+}$ in the 
             ${\rm K^-\pi^+\pi^+}$ channel (right), 
             in 0--5\% central \mbox{Pb--Pb} collisions 
             and in pp collisions.} 
    \label{fig:Destat}
  \end{center}
\end{figure}

{\it Exclusive charm meson reconstruction}.
Among the most promising channels for open charm detection are the 
$\rm D^0 \to K^-\pi^+$ ($c\tau\approx 120~\mum$, branching ratio 
$\approx 3.8\%$) and $\rm D^+ \to K^-\pi^+\pi^+$ ($c\tau\approx 300~\mum$, 
branching ratio $\approx 9.2\%$) decays. The detection strategy
to cope with the large combinatorial background from the underlying event 
is based on the selection of displaced-vertex topologies, i.e. separation 
from the primary vertex of
the tracks from the secondary vertex 
and good alignment between the reconstructed D meson momentum 
and flight-line~\cite{alicePPR2,elena}. 
An invariant-mass analysis is used to extract the raw signal 
yield, to be then corrected for selection and reconstruction efficiency
and for detector acceptance.
As shown in Fig.~\ref{fig:Destat} (left),
the accessible $\pt$ range for the $\Dz$ is $1$--$20~\gev/c$ in \mbox{Pb--Pb} and 
$0.5$--$20~\gev/c$ in pp, 
with statistical errors better than 15--20\% at high $\pt$. Similar capability 
is expected for the $\rm D^+$ (right-hand panel), 
though at present the
statistical errors are estimated only in the range $1<\pt<8~\gev/c$.
The systematic errors 
(acceptance and efficiency corrections, 
centrality selection for Pb--Pb) are expected to be smaller than 20\%.

{\it Beauty via single electrons.}
The main sources of background electrons are: decays of D mesons; 
$\pi^0$ Dalitz decays 
and decays of light vector mesons (e.g.\,$\rho$ and $\omega$);
conversions of photons in the beam pipe or in the inner detector 
layer and pions misidentified as electrons. 
Given that electrons from beauty have average 
impact parameter $d_0\simeq 500~\mum$
and a hard $\pt$ spectrum, it is possible to 
obtain a high-purity sample with a strategy that relies on:
electron identification with a combined $\dEdx$ (TPC) and transition
radiation (TRD) selection;
impact parameter cut to 
reduce the charm-decay component and 
reject misidentified $\pi^\pm$ and $\rm e^{\pm}$
from Dalitz decays and $\gamma$ conversions.
As an example, with $d_0>200~\mum$ and $\pt>2~\gev/c$, the expected statistics
of electrons from b decays is $8\times 10^4$ for $10^7$ central 
\mbox{Pb--Pb} events, allowing the measurement of electron-level 
$\pt$-dif\-fe\-ren\-tial 
cross section in the range $2<\pt<20~\gev/c$ with statistical errors smaller
than 15\% at high $\pt$. Similar performance figures are expected for 
pp collisions~\cite{julien}.

{\it Beauty via muons.}
B production in \mbox{Pb--Pb} collisions 
can be measured also in the ALICE muon 
spectrometer ($-4<\eta<-2.5$) analyzing the single-muon $\pt$ 
distribution~\cite{alicePPR2}.
The main backgrounds to the `beauty muon' signal are $\pi^\pm$, 
$\rm K^\pm$ and charm decays. The cut $\pt>1.5~\gev/c$ is applied to all
reconstructed muons in order to increase the signal-to-background ratio.
Then, a fit technique allows to extract a $\pt$ distribution of muons 
from B decays.
Since only minimal cuts are applied, the statistical errors are 
expected to be smaller than 5\% up to muon 
$\pt\approx 30~\gev/c$~\cite{julien}.

\begin{figure}[!t]
  \begin{center}
    \includegraphics[width=0.52\textwidth]{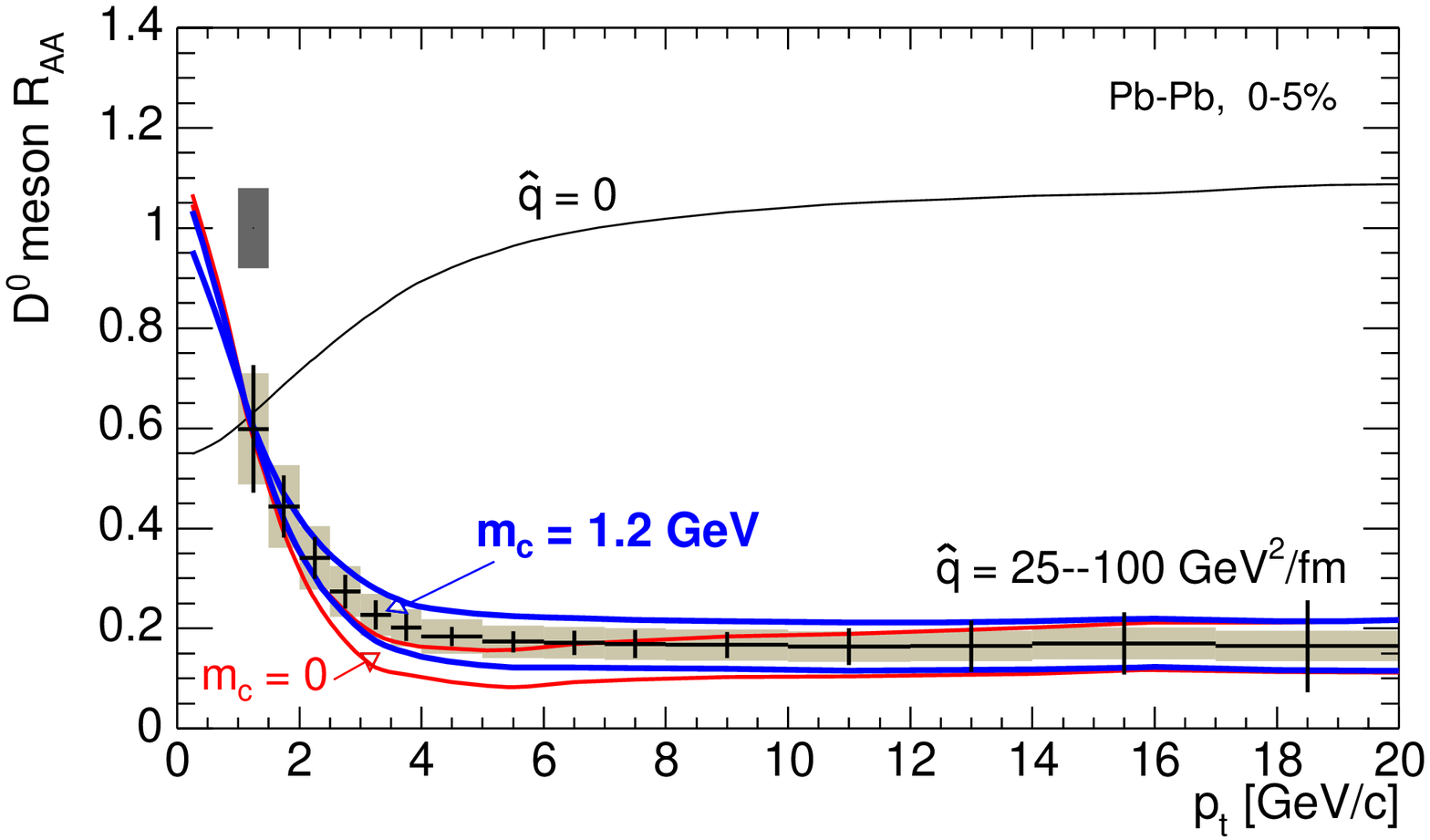}
    \includegraphics[width=0.47\textwidth]{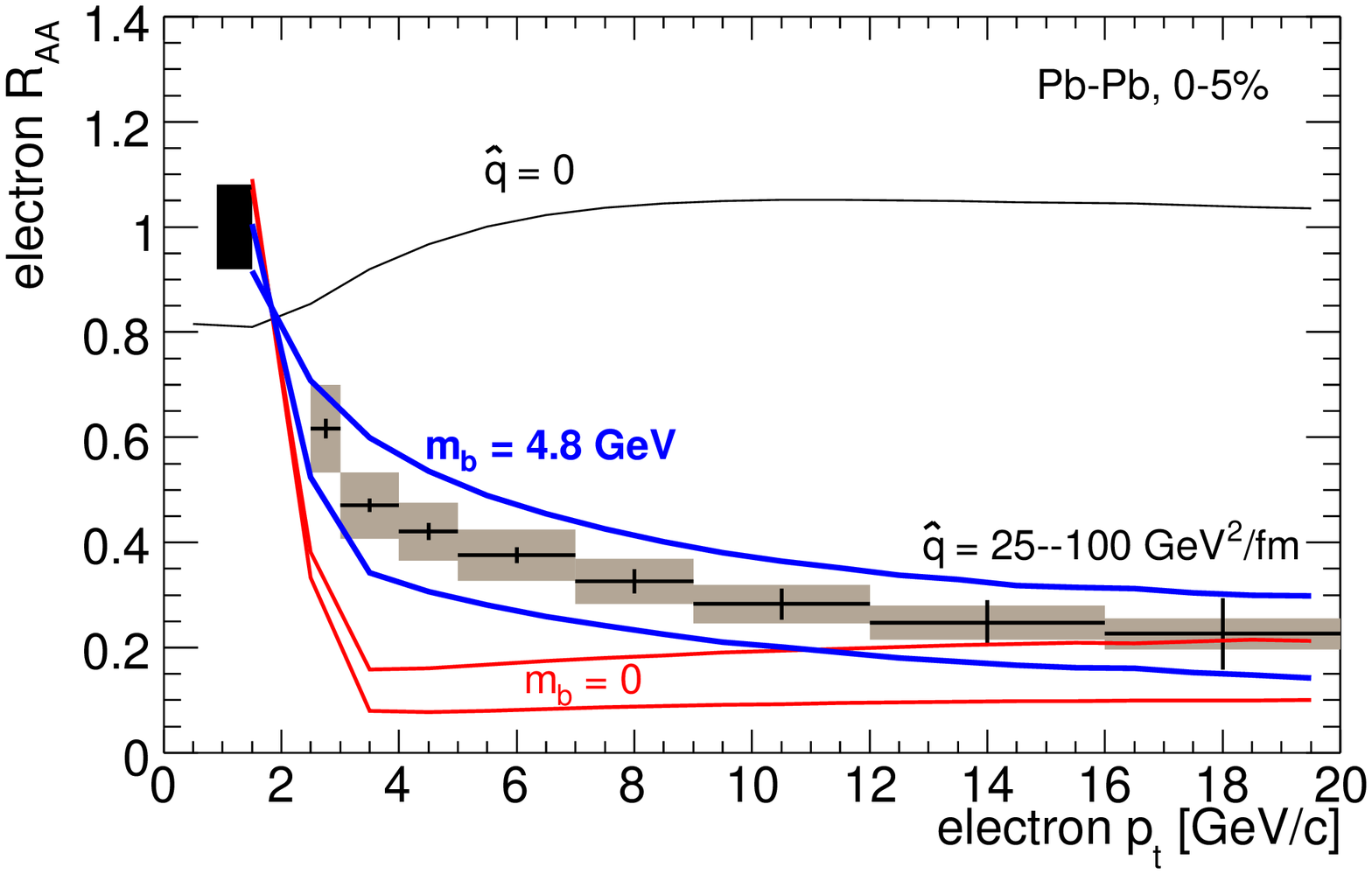}
    \caption{Nuclear modification factors for $\Dz$ mesons 
             (left) and for B-decay electrons
             (right).
             Errors corresponding to the centre of the prediction bands   
             for massive quarks are shown: bars = statistical, 
             shaded area = systematic.} 
    \label{fig:RAA}
  \end{center}
\end{figure}

{\it Nuclear modification factors.}
We investigated the possibility of using 
the described charm and beauty measurements 
to study the high-$\pt$ suppression induced by parton energy loss.
The sensitivity to $\RAA^{\rm D}$ 
and $\RAA^{\rm e~from~B}$ is presented in Fig.~\ref{fig:RAA}.
Predictions~\cite{adsw} with and without the effect of the
heavy-quark mass, for a medium transport coefficient $\hat{q}$ 
(a measurement of the medium density) in the range 
$25$--$100~\gev^2/\fm$, are also shown.

\section{Quarkonia capabilities}
\label{quarkonia}

\begin{figure}[!t]
  \begin{center}
    \includegraphics[width=0.75\textwidth]{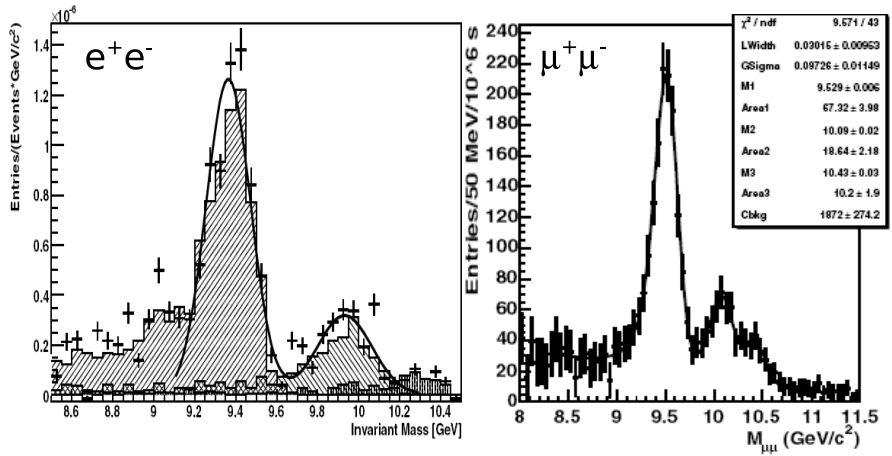}
    \caption{The signal of $\Upsilon$ states in central \mbox{Pb--Pb} 
             collisions, 
             as reconstructed by ALICE~\cite{alicePPR2}, in the dielectron
	     and in the dimuon channel, 
	     in one month of data-taking.} 
    \label{fig:upsilon}
  \end{center}
\end{figure}

ALICE can detect quarkonia 
in the dielectron channel at central rapidity ($|y|\lsim 1$) and in the
dimuon channel at forward rapidity ($-4\lsim y\lsim -2.5$). In both channels
the acceptance extends down to zero transverse momentum, since the 
minimum $\pt$ for e and $\mu$ identification is about $1~\gev/c$. 
The high $\pt$ reach 
is expected to be 10~(20)~$\gev/c$ for the J/$\psi$ in 
$\rm e^+e^-$ ($\mu^+\mu^-$), for a \mbox{Pb--Pb} 
run of one month at nominal luminosity.
We emphasized the importance of separating the $\Upsilon$ and 
$\Upsilon^\prime$, to probe the initial temperature of the medium;
given that the mass difference between the two bottomonium states is 
about $500~\mev$, a mass resolution of order $100~\mev$ at 
$M_{\ell^+\ell^-}\sim 10~\gev$, i.e. $\sigma_M/M\approx 1\%$, is required.
This requirement is fulfilled for both dielectrons and dimuons,
with a mass resolution of about $90~\mev$.
For illustration, in Fig.~\ref{fig:upsilon} 
we show the simulated dilepton mass spectra in the 
$\Upsilon$ region after background subtraction~\cite{alicePPR2}.
 
Simulation studies are in progress to prepare 
a measurement of the fraction of J/$\psi$
that feed-down from B decays. Such measurement can be performed 
by studying the separation of the dilepton pairs in the J/$\psi$ 
invariant mass region
from the main interaction vertex. The analysis is also expected to provide a 
measurement of the beauty $\pt$-differential cross section.

\section{Summary}
\label{summary}

We have discussed how heavy quarks, abundantly produced at LHC energies, 
will allow to address several issues at the heart of 
in heavy-ion physics. 
They provide tools to probe the density
(via parton energy loss and its predicted mass dependence)
and the temperature
(via successive dissociation patterns of quarkonia)
of the high-density QCD medium formed in \mbox{Pb--Pb} collisions.
The excellent tracking, vertexing and particle identification performance 
of ALICE will allow to fully explore this rich phenomenology.

\vspace{.4cm}

\end{document}